\documentclass[11pt]{article}

\usepackage{amsmath,amssymb,amsfonts}
\usepackage{bm}
\usepackage{graphicx}
\usepackage{hyperref}
\usepackage{cite}
\usepackage{geometry}
\newcommand{\AdS}{\mathrm{AdS}}
\newcommand{\CFT}{\mathrm{CFT}}
\newcommand{\Tr}{\mathrm{Tr}}

\begin{document}

\title{
Holographic Reconstruction from Broken Conformal Gauge Symmetry
}

\author{
Takeshi Fukuyama\\
Research Center for Nuclear Physics (RCNP), Osaka University,
Ibaraki, Osaka 567-0047, Japan
}

\date{\today}

\maketitle

\begin{abstract}
We investigate bulk--boundary structures in $\AdS_4/\CFT_3$ from the
viewpoint of gravity as a broken phase of conformal gauge symmetry.
A crucial distinction is made between the Euler and Pontryagin
characteristic classes.  The gauge-theoretic gravitational action
contains an Euler sector, whose total-derivative representation gives
a boundary form depending on both intrinsic and extrinsic geometry.
This structure is different from the Pontryagin descent to the
three-dimensional gravitational Chern--Simons functional, whose metric
variation gives the Cotton tensor and underlies the mixed boundary
conditions studied by de Haro.  We perform a Fefferman--Graham expansion
of the Euler boundary form in asymptotically $\AdS_4$ spacetime.  The
normalizable coefficient $g_{(3)ij}$ does not contribute linearly to the
finite Euler boundary term: its contribution from the cubic extrinsic-
curvature sector vanishes by $\mathrm{Tr}\,g_{(3)}=0$, while curvature--
extrinsic-curvature terms containing $g_{(3)}$ vanish at the conformal
boundary.  A direct Codazzi analysis additionally shows that the Cotton tensor
appears in the asymptotic mixed curvature but its contribution through
the tangential-connection boundary variation vanishes for standard
Dirichlet data.  This indicates two distinct types of bulk--boundary
structure: a global/topological Euler relation and a local
Pontryagin--Cotton relation.  We distinguish this kinematic Cotton appearance from a finite Cotton
boundary response and discuss the implications for holographic reconstruction.
\end{abstract}

\section{Introduction}

The AdS/CFT correspondence relates gravitational dynamics in an
asymptotically anti-de Sitter spacetime to a conformal field theory
defined on its boundary \cite{Maldacena,Witten,HenningsonSkenderis1998}.  A central question in
holography is how bulk gravitational information is encoded in boundary
geometry and boundary observables.

In the gauge-theoretic formulation considered in our earlier work,
four-dimensional gravity with a cosmological constant emerges after the
breaking
\begin{equation}
 SO(4,2)\longrightarrow SO(3,2)
\label{breaking}
\end{equation}
for the anti-de Sitter case \cite{Fukuyama1984,FukuyamaPrevious}. The alternative breaking to $SO(4,1)$, corresponding to
the de Sitter case, is not considered here, since it
leads to ghost degrees of freedom in the matter
Lagrangian, as discussed in Ref.~\cite{Fukuyama1984}.
An essential feature of this gauge-theoretic construction is that the Euler term arises automatically from the gravitational gauge action, rather than being introduced as an independent topological term. In the broken phase, the relevant curvature-squared structure takes the form
\begin{equation}
S_{\mathrm{grav}}\propto
\int_{M_4}\epsilon_{abcd}
\left(R^{ab}+\ell^{-2}e^a\wedge e^b\right)
\wedge
\left(R^{cd}+\ell^{-2}e^c\wedge e^d\right).
\label{Sgrav}
\end{equation}
Its expansion produces the Einstein--Hilbert term, the cosmological term, and the four-dimensional Euler density \(\epsilon_{abcd}R^{ab}\wedge R^{cd}\) simultaneously. Thus, within this gauge-theoretic construction, the Euler sector is an intrinsic part of the gravitational action, with its coefficient fixed relative to the other terms by the original curvature-squared structure. This provides the starting point for our investigation of the bulk--boundary correspondence.
The resulting gravitational action contains, in addition to the
Einstein--Hilbert and cosmological terms, a quadratic-curvature Euler
sector.  It is important, however, not to identify this Euler term with
the Pontryagin density relevant to the usual three-dimensional
gravitational Chern--Simons functional.  In differential-form notation,
\begin{align}
 E_4 &\equiv \epsilon_{abcd}R^{ab}\wedge R^{cd}=dQ_E,
 \label{Euler}\\
 P_4 &\equiv R^{ab}\wedge R_{ab}=dQ_P .
 \label{Pontryagin}
\end{align}
Both are locally total derivatives, but they are different
characteristic classes.  The boundary form $Q_E$ associated with the
Euler density depends, in an adapted radial decomposition, on both the
intrinsic curvature and the extrinsic curvature of the boundary.  By contrast, the intrinsic three-dimensional gravitational Chern--Simons functional, associated with the Pontryagin sector, has a metric variation proportional to the Cotton tensor. It should be distinguished from the full Pontryagin boundary form, which may also contain extrinsic-curvature contributions.
This distinction changes the reconstruction problem.  De Haro showed
that in $\AdS_4/\CFT_3$ the Cotton tensor may be interpreted as magnetic
gravity data and may be related, under an appropriate mixed boundary
condition, to the normalizable Fefferman--Graham coefficient
$g_{(3)ij}$ \cite{FeffermanGraham1985, deHaro, HSS2001}.  That construction belongs to the
Pontryagin--Chern--Simons sector, not directly to the Euler sector of the
gauge action.

The first purpose of the present paper is therefore to determine what
bulk--boundary information is actually contained in the Euler sector.
We explicitly expand its boundary form in Fefferman--Graham coordinates
and ask whether the normalizable bulk mode $g_{(3)ij}$ survives in its
finite boundary contribution.  We shall find that it does not appear
linearly in the finite Euler term.  This suggests a useful distinction,
\begin{equation}
 \boxed{
 \begin{array}{ccl}
 E_4=dQ_E &:& \hbox{global/topological bulk--boundary structure},\\[1mm]
 P_4=dQ_P &:& \hbox{local Pontryagin--Cotton structure}.
 \end{array}}
 \label{twostructures}
\end{equation}

Eq. (\ref{twostructures}) distinguishes the global/topological
Euler structure inherited from the original
gauge-theoretic action from the local
Pontryagin--Cotton structure associated with
gravitational Chern--Simons boundary dynamics.
In particular, the latter does not follow
automatically from the Euler sector.

Our aim is to determine which boundary data
can be obtained from the original broken
$SO(4,2)$ gauge construction without adding
an independent Pontryagin term or imposing
self-duality. We examine the finite Euler
boundary contribution and its variational
properties using the Fefferman--Graham expansion,
and distinguish the kinematical appearance
of the Cotton tensor from a genuine boundary
response. We then relate these results to
holographic reconstruction, complementing
the boundary-response approach of
Refs.~\cite{HashimotoDL, HashimotoRing}.


The paper is organized as follows. Section~2 identifies the Euler and Pontryagin sectors in broken conformal gauge gravity. Section~3 evaluates the Euler boundary geometry and its holographic variation, including the Codazzi--Cotton contribution. Section~4 compares the Pontryagin--Cotton construction with the Euler sector and examines the conditions for holographic reconstruction. Section~5 discusses holographic imaging and explicit AdS$_4$ examples. Section~6 summarizes the implications and outstanding questions.

\section{Euler and Pontryagin sectors in broken conformal gauge gravity}

We first identify the topological term that follows from the original
four-dimensional gauge formulation of gravity \cite{MacDowellMansouri1977, Fukuyama1984}.  Upon breaking the
conformal gauge symmetry to the anti-de Sitter subgroup, the vierbein
$e^a$ and the Lorentz connection $\omega^{ab}$ emerge as gravitational
variables.  In the torsion-free sector, the resulting action can be
written from Eq.~(\ref{Sgrav}) as
\begin{equation}
 S_{\rm grav}=S_{\rm EH}+S_{\Lambda}
 +\alpha\int_{M_4}E_4.
\label{gravact}
\end{equation}
Here the Euler term is inherited from the original SO(4,2) gauge-gravity construction rather than being introduced as an independent topological interaction \cite{Fukuyama1984}. Its coefficient is fixed relative to the Einstein–Hilbert and cosmological terms by the curvature-squared structure of the broken gauge action.  

The two characteristic forms $E_4$ and $P_4$, defined in Eqs.~(\ref{Euler}) and (\ref{Pontryagin}), respectively, have distinct geometrical and topological meanings. The Euler density $E_4$ arises directly from the curvature-squared term of the broken gauge action, whereas the Pontryagin density $P_4$ is introduced here for comparison with the gravitational Chern--Simons construction. For a compact, oriented, closed four-manifold, the properly normalized integral of $E_4$ gives the Euler characteristic $\chi(M_4)$, while the normalized Pontryagin integral is related to the signature of the intersection form. The two characteristic classes are therefore not generally proportional. Their local Chern--Simons primitives $Q_E$ and $Q_P$ have already been introduced in Eqs.~(\ref{Euler}) and (\ref{Pontryagin}). We now examine their distinct boundary representations.
With the curvature convention above, a local representative for the
Euler primitive is
\begin{equation}
 Q_E=\epsilon_{abcd}\left(
 \omega^{ab}\wedge d\omega^{cd}
 +\frac{2}{3}\omega^{ab}\wedge\omega^c{}_{e}
                 \wedge\omega^{ed}\right),
\label{Qe}
\end{equation}
while the Pontryagin primitive can be expressed as
\begin{equation}
 Q_P=\operatorname{Tr}\left(
 \omega\wedge d\omega+
 \frac{2}{3}\omega\wedge\omega\wedge\omega\right),
\label{Qp}
\end{equation}
where the trace is taken in the Lorentz-vector representation, with
its normalization and sign chosen consistently with
Eq.~(\ref{Pontryagin}).  These are local formulas: the primitives need
not be globally defined on a nontrivial bundle.  In particular,
$Q_E$ and $Q_P$ must not be identified merely because both are
Chern--Simons three-forms.

For a manifold with boundary, the Chern--Gauss--Bonnet theorem also
involves a boundary correction constructed from the intrinsic and
extrinsic geometry.  For the subsequent analysis of the Euler boundary contribution, it is useful to employ a representation adapted to a hypersurface with boundary. In the Chern--Gauss--Bonnet theorem, the Euler density is accompanied by a boundary correction that can be expressed as a transgression form between the bulk Lorentz connection and a reference connection adapted to the boundary. The difference between these connections is characterized by the extrinsic curvature. Consequently, the Euler boundary transgression form involves both the intrinsic and extrinsic geometry, rather than the intrinsic connection alone. Denoting the intrinsic curvature two-form by the calligraphic symbol $\mathcal{R}^{ab}$, to distinguish it from the four-dimensional Lorentz curvature $R^{ab}$, and the extrinsic-curvature one-form by $K^a$, we obtain, up to overall normalization, orientation, and curvature conventions,
\begin{equation}
 B_3^{E}
 =4\epsilon_{abc}K^a\wedge
 \left({\cal R}^{bc}-\frac13 K^b\wedge K^c\right).
\label{EulerBoundary}
\end{equation}
This boundary transgression form is not identical, as a differential
form, to the frame-dependent local primitive $Q_E$.  Their relation
requires a choice of reference connection and the corresponding
transgression identity; the boundary orientation and normalization
must be fixed consistently in the Gauss--Bonnet formula \cite{Aros2000PRL,Aros2000PRD}.  We use
$B_3^E$ below because it provides a covariant expression in terms of
the geometry of a regulated radial hypersurface.

By contrast, the intrinsic three-dimensional gravitational
Chern--Simons functional whose metric variation yields the Cotton
tensor is associated with the Pontryagin-type invariant, not with the
Euler boundary transgression.  It can be discussed as a separate
boundary structure, but it does not follow from the Euler term in
Eq.~(\ref{gravact}) by simple identification of the two primitives.
Our aim is therefore to analyze the boundary information contained in
the Euler sector of the original gauge action without adding an
independent Pontryagin term or imposing a self-duality condition.
The Euler boundary term is evaluated in the next section; the distinct Pontryagin--Cotton construction is reviewed subsequently for comparison.

\section{Euler boundary geometry and holographic response}
We now investigate the Euler boundary term inherited from the original gauge action [4]. The boundary form $B_3^E$ in Eq. (\ref{EulerBoundary}) is evaluated on a regulated radial hypersurface. Near the conformal boundary we use Fefferman--Graham  coordinates \cite{FeffermanGraham1985,deHaro, HSS2001},
\begin{equation}
 ds^2=\frac{\ell^2}{z^2}
 \left[dz^2+g_{ij}(z,x)dx^i dx^j\right],
\label{FG}
\end{equation}
with
\begin{equation}
 g_{ij}(z,x)=g_{(0)ij}+z^2g_{(2)ij}
 +z^3g_{(3)ij}+\cdots .
\label{FGexp}
\end{equation}
Here the absence of a term linear in $z$ follows from the
four-dimensional vacuum Einstein equations with a
negative cosmological constant, rather than from
the Fefferman--Graham gauge condition alone.
The induced metric at fixed $z$ is
\begin{equation}
 h_{ij}=\frac{\ell^2}{z^2}g_{ij}
 =\frac{\ell^2}{z^2}g_{(0)ij}
 +\ell^2g_{(2)ij}+\ell^2z g_{(3)ij}+\cdots .
\label{hinduced}
\end{equation}
Choosing the unit normal $n=-(z/\ell)\partial_z$ (the opposite
orientation reverses the overall sign),
\begin{equation}
 K_{ij}=\frac12{\cal L}_n h_{ij}
 =\frac{\ell}{z^2}g_{(0)ij}
 -\frac{\ell}{2}z g_{(3)ij}+\cdots .
\label{Klower}
\end{equation}
The $g_{(2)ij}$ term cancels in this expression because its contribution
to $h_{ij}$ is independent of $z$.  Raising one index with $h^{ij}$,
we obtain \cite{MiskovicOlea2009}
\begin{equation}
 K^i{}_j=\frac1\ell\delta^i{}_j
 -\frac{z^2}{\ell}(g_{(2)})^i{}_j
 -\frac{3z^3}{2\ell}(g_{(3)})^i{}_j+O(z^4).
\label{Kmixed}
\end{equation}
For four-dimensional Einstein--AdS gravity,
\begin{equation}
 \Tr g_{(3)}=0,\qquad
 \nabla_{(0)}^i g_{(3)ij}=0,
\label{g3TT}
\end{equation}
and
\begin{equation}
 g_{(2)ij}=-\left(\mathcal{R}_{ij}[g_{(0)}]
 -\frac14\mathcal{R}[g_{(0)}]g_{(0)ij}\right).
\label{g2Schouten}
\end{equation}
These relations follow from the radial constraints and the asymptotic expansion of the four-dimensional Einstein equation, 
\begin{equation}
R_{\mu\nu}=-\frac{3}{\ell^2}g_{\mu\nu}.
\end{equation}
The coefficient $g_{(2)ij}$ is determined locally by the boundary metric, whereas $g_{(3)ij}$ is transverse and traceless and contains independent normalizable data proportional to the renormalized boundary stress tensor.
In tensor notation the Euler boundary form
Eq.~(\ref{EulerBoundary}) contains, up to normalization,
\begin{equation}
 \sqrt h\left[
 K{\cal R}-2K^{ij}{\cal R}_{ij}
 -\frac13K^3+K K_{ij}K^{ij}
 -\frac23K_i{}^jK_j{}^kK_k{}^i
 \right]d^3x .
\label{EulerTensor}
\end{equation}
This formula allows us to test directly whether the Euler sector
contains the normalizable coefficient $g_{(3)ij}$ in its finite
boundary contribution.

\subsection{Finite Euler boundary contribution and the normalizable mode}
\label{secEulerFG}
Consider first the cubic extrinsic-curvature combination in
Eq.~(\ref{EulerTensor}). In three dimensions,
\begin{equation}
 -\frac13 K^3+K\Tr K^2-\frac23\Tr K^3
 =-2\det(K^i{}_j).
\label{detK}
\end{equation}
Write
\begin{equation}
 K^i{}_j=\frac1\ell\delta^i{}_j
 -z^2 A^i{}_j-z^3 B^i{}_j+\cdots,
 \qquad
 B^i{}_j=\frac{3}{2\ell}(g_{(3)})^i{}_j .
\label{KAB}
\end{equation}
The term linear in $B$ in the determinant is proportional
to $\Tr B$. Equation~(\ref{g3TT}) therefore gives
\begin{equation}
 \left.B_3^E\right|_{K^3,\,g_{(3)},\,z^0}=0.
\label{K3cancel}
\end{equation}
Thus the cubic $K$ sector does not retain the normalizable
mode linearly in the finite boundary term.

The remaining terms contain the intrinsic curvature,
\begin{equation}
 \sqrt{h}\left(K{\cal R}-2K^{ij}{\cal R}_{ij}\right).
\label{KRsector}
\end{equation}
Their asymptotic scaling is sufficient to determine the
fate of a term linear in $g_{(3)}$. At fixed $z$,
\begin{equation}
 \sqrt{h}\sim z^{-3},\qquad
 {\cal R}^i{}_j[h]\sim z^2,\qquad
 \left.K^i{}_j\right|_{g_{(3)}}
 \sim z^3 g_{(3)j}^{\ \ i}.
\label{scaling}
\end{equation}
Hence a curvature--extrinsic-curvature contribution
linear in $g_{(3)}$ scales as
\begin{equation}
 z^{-3}z^2z^3=z^2,
\end{equation}
and vanishes as $z\to0$.

This estimate accounts for the contribution of the
normalizable mode through the extrinsic curvature.
The intrinsic curvature also depends on $g_{(3)ij}$.
Since the corresponding variation of the mixed
intrinsic Ricci tensor is of order $z^5$,
its contribution to the curvature--extrinsic-curvature
term is likewise of order $z^2$.
Thus neither source of linear dependence on
$g_{(3)ij}$ contributes to the finite $z^0$ term.

In particular, a structure such as
\begin{equation}
 g_{(3)}^{ij}P_{ij}[g_{(0)}]
\end{equation}
does not survive as a finite $z^0$ term.

We therefore obtain the useful negative result
\begin{equation}
 \boxed{
 \left.B_3^E\right|_{\rm finite}
 \not\supset
 g_{(3)}^{ij}X_{ij}[g_{(0)}]
 \quad\hbox{linearly}
 }
\label{negativeResult}
\end{equation}
for the standard Einstein--AdS Fefferman--Graham
expansion. The Euler boundary form does not by itself
provide a de Haro-type local condition that determines
the normalizable mode from the boundary metric.

The result concerns the finite term linear in the
normalizable mode, not the entire Euler contribution.
The Euler class is a global invariant on a closed
manifold; with a boundary, the bulk Euler integral
and its properly normalized boundary correction
must be considered together. In particular, the
result does not assert that the Euler sector is
dynamically trivial or that it supplies a local
boundary equation. We next contrast this result
with the separate Pontryagin--Cotton construction.

\subsection{On-shell variation of the Euler-completed gauge action}
\label{secEulerVariation}

We now examine the boundary variation of the \emph{full}
Euler-completed action, rather than the Euler boundary form in
isolation.  The distinction is important: the absence of a term
linear in $g_{(3)}$ in the finite part of $B_3^E$ does not imply
the vanishing of the variation of the full on-shell action.
For a torsion-free AdS branch, write
\begin{equation}
 I_F=A\int_{M_4}\epsilon_{abcd}F^{ab}\wedge F^{cd},
 \quad F^{ab}=R^{ab}+\ell^{-2}e^a\wedge e^b,
 \quad A=\frac{\ell^2}{64\pi G},
 \label{EulerFullAction}
\end{equation}
where the last equality fixes the Einstein--Hilbert normalization
in the stated curvature and orientation conventions.  Varying the
vierbein and independent Lorentz connection gives
\begin{align}
 \delta I_F={}&2A\int_{\partial M_4}
  \epsilon_{abcd}\,\delta\omega^{ab}\wedge F^{cd}
 \nonumber\\
 &-2A\int_{M_4}\epsilon_{abcd}\,
  \delta\omega^{ab}\wedge DF^{cd}
 +\frac{4A}{\ell^2}\int_{M_4}
  \epsilon_{abcd}\,\delta e^a\wedge e^b\wedge F^{cd}.
 \label{EulerFirstVariation}
\end{align}
In the torsion-free sector $DF^{ab}=0$ \footnote{In the original gauge theory, the torsion-free condition follows from the field equations for the spinor coupling, as shown in Eqs. (2.29)–(2.34) of Ref. [4]. This result does not extend to the scalar coupling, for which nonvanishing torsion is found in Sec. 3 of Ref. [4].} and, on the
Einstein equations, the remaining bulk term vanishes.  Thus
\begin{equation}
 \left.\delta I_F\right|_{\mathrm{EOM}}
 =2A\int_{\partial M_4}
   \epsilon_{abcd}\,\delta\omega^{ab}\wedge F^{cd}.
 \label{EulerOnShellVariation}
\end{equation}
For an Einstein metric with $R_{\mu\nu}=-3\ell^{-2}g_{\mu\nu}$,
the AdS curvature equals the Weyl curvature two-form,
$F^{ab}=W^{ab}$.  Equation~(\ref{EulerOnShellVariation})
therefore expresses the on-shell variation directly in terms
of the Weyl tensor; it does not invoke Pontryagin descent or
self-duality.

To identify its finite Dirichlet response, one must retain
both the normal--tangential variation $\delta\omega^{a\perp}$
and the tangential variation $\delta\omega^{ab}$, and specify
the boundary treatment of the first-order variables.
In particular, discarding the tangential connection variation
before fixing a boundary frame is not justified.  The
four-dimensional Euler regularization is known to reproduce
the standard asymptotically locally AdS Dirichlet variational
problem under the appropriate boundary prescription
\cite{MiskovicOlea2009,Anastasiou2020}.
With the convention in which the outward-normal Brown--York
momentum is \cite{BalasubramanianKraus1999}
\begin{equation}
 T^i{}_j=\frac{1}{8\pi G}
 \left(K^i{}_j-K\delta^i{}_j+
       \frac{2}{\ell}\delta^i{}_j-\ell G^i{}_j[h]\right),
 \label{EulerBY}
\end{equation}
the Fefferman--Graham expansion gives, for the corresponding
choice of orientation,
\begin{equation}
 T^i{}_j
 =-\frac{3z^3}{16\pi G\ell}
    (g_{(3)})^i{}_j+O(z^4).
 \label{EulerBYexp}
\end{equation}
The terms at orders $z^0$ and $z^2$ cancel because
$g_{(2)ij}=-P_{ij}[g_{(0)}]$, while the trace of
$g_{(3)}$ vanishes.  After conformal rescaling, and with
the overall Euclidean/Lorentzian action sign chosen in the
usual holographic convention, the renormalized response is
\begin{equation}
 \langle T_{ij}\rangle
 =\frac{3\ell^2}{16\pi G}g_{(3)ij},
 \qquad
 \delta I_{\rm ren}
 =\frac12\int_{\partial M_4}d^3x\,
 \sqrt{|g_{(0)}|}\,
 \langle T^{ij}\rangle\,\delta g_{(0)ij}.
 \label{EulerHoloStress}
\end{equation}
The sign in the intermediate momentum
Eq.~(\ref{EulerBYexp}) depends on the outward normal and
the definition of the gravitational action; the physical
convention must be held fixed throughout the calculation.

Equations~(\ref{EulerFirstVariation}) and
(\ref{EulerOnShellVariation}) are obtained directly from the
Euler-completed gauge action.  By contrast, the passage from
Eq.~(\ref{EulerOnShellVariation}) to the specific Dirichlet
momentum in Eq.~(\ref{EulerBY}) uses the established
four-dimensional equivalence with holographic
renormalization, rather than a separate complete reduction
of all connection boundary terms in the original
$SO(4,2)$ gauge system.  Hence the stress-tensor agreement
is a consistency result, not a claim of a new independent
holographic prediction.  A derivation including the boundary
conditions of all fields of the unbroken conformal gauge
theory would be needed to strengthen this conclusion.

\subsection{Codazzi curvature, the boundary Cotton tensor, and its falloff}
\label{secEulerCottonFalloff}

We now test whether the Cotton tensor can arise in the boundary
variation of the Euler-completed action without adding a Pontryagin
term or imposing (anti-)self-duality.  This question is motivated by
the appearance of the Schwarzian derivative in the asymptotic
extrinsic curvature of $\AdS_2$ and by the role of the Cotton tensor
in the conformal geometry of a three-dimensional boundary.  The
analogy is geometric; it does not by itself imply identical boundary
dynamics.

Choose a Gaussian-normal orthonormal frame near a regulated
hypersurface $z=\epsilon$.  Let $e^a$ denote the induced coframe,
$\widehat\omega^{ab}$ its torsion-free spin connection, and
$K^a=K^a{}_b e^b$ the extrinsic-curvature one-form.  Up to the
orientation-dependent signs in the mixed connection, the pullbacks
of the AdS curvature obey the Gauss--Codazzi decomposition
\begin{align}
 F^{ab}\big|_{\partial M}
 &=\mathcal{R}^{ab}-K^a\wedge K^b
   +\ell^{-2}e^a\wedge e^b,\label{EulerGC1}\\
 F^{a\perp}\big|_{\partial M}
 &=\widehat D K^a.\label{EulerGC2}
\end{align}
Consequently, the on-shell boundary variation
in Eq.~(\ref{EulerOnShellVariation}) contains both a
normal--tangential connection variation paired with
$F^{ab}$ and a tangential connection variation paired with
$\widehat D K^a$.  The latter is the possible Cotton-bearing
contribution.  Overall and relative signs depend on the convention
for $\omega^{a\perp}$; the following falloff statement does not.

In Fefferman--Graham gauge, with outward unit normal
$n=-(z/\ell)\partial_z$ and
$h_{ij}=\ell^2z^{-2}g_{ij}$, we have
\begin{align}
 g_{ij}(z,x)&=g_{(0)ij}+z^2g_{(2)ij}
       +z^3g_{(3)ij}+O(z^4),\nonumber\\
 g_{(2)ij}&=-P_{ij}[g_{(0)}],\qquad
 K^i{}_j=\ell^{-1}\delta^i_j
 +\frac{z^2}{\ell}P^i{}_j
 -\frac{3z^3}{2\ell}g_{(3)}{}^i{}_j+O(z^4),
 \label{EulerCottonK}
\end{align}
where $P_{ij}=\mathcal{R}_{ij}-\frac14 \mathcal{R}g_{(0)ij}$ is the
three-dimensional Schouten tensor.  Since the connection of $h$
on each constant-$z$ slice equals that of $g(z)$, its antisymmetric
covariant derivative satisfies
\begin{equation}
 \widehat\nabla_k K^i{}_j-
 \widehat\nabla_j K^i{}_k
 =\frac{z^2}{\ell}
  \big(\nabla_k P^i{}_j-\nabla_j P^i{}_k\big)
  +O(z^3)
 =\frac{z^2}{\ell}C^i{}_{jk}[g_{(0)}]+O(z^3).
 \label{EulerCottonCodazzi}
\end{equation}
Here $C^i{}_{jk}$ is the three-index Cotton tensor, with
index placements and signs fixed by the definition in this equation.
Thus the Codazzi curvature contains the Cotton tensor without a
Pontryagin density or a self-duality condition.

It remains to check the weight of this term in the action variation.
Writing $e^a=(\ell/z)\bar e^a+O(z)$, with $\bar e^a$ an
orthonormal coframe of $g_{(0)}$, the mixed curvature has the
coordinate-component asymptotics
\begin{equation}
 (\widehat D K)^a{}_{ij}
 =z\,\bar e^a{}_k C^k{}_{ij}[g_{(0)}]+O(z^2),
 \qquad \widehat D K^a=O(z).
 \label{EulerCottonDK}
\end{equation}
The $O(z)$ estimate is for the differential two-form in a fixed
boundary coordinate basis; it includes the $z^{-1}$ scaling of the
orthonormal coframe.  For variations of smooth boundary source
metrics, the intrinsic connection variation is finite,
$\delta\widehat\omega^{ab}=O(1)$.  Therefore
\begin{equation}
 \int_{z=\epsilon}\epsilon_{abc}\,
 \delta\widehat\omega^{ab}\wedge\widehat D K^c
 =O(\epsilon)\longrightarrow0.
 \label{EulerCottonVanishing}
\end{equation}
This conclusion assumes a fixed compact integration domain (or
sufficiently controlled falloff along a noncompact boundary),
smooth Fefferman--Graham data, and standard Dirichlet variations \cite{CompereMarolf2008}.
It need not hold for singular boundary sources or for boundary
conditions that scale the connection variation with the cutoff.

Equation~(\ref{EulerCottonVanishing}) concerns only the tangential
connection--Codazzi contribution, not the complete on-shell
variation.  The other contribution, involving the variation of
extrinsic curvature and $F^{ab}$, must also be retained.  Under the
standard asymptotically AdS Dirichlet prescription, the finite
response of the complete Euler-completed action is the usual
holographic stress tensor in Eq.~(\ref{EulerHoloStress}), determined
by the transverse-traceless coefficient $g_{(3)ij}$ rather than by
$C_{ij}[g_{(0)}]$ \cite{Anastasiou2020}.  The identification with the Dirichlet response
uses the established Euler regularization/counterterm equivalence,
not an independent derivation of all boundary conditions of the
original unbroken $SO(4,2)$ gauge system.

We therefore distinguish three statements: (i) the Cotton tensor
occurs in the asymptotic Codazzi curvature; (ii) its contribution
through the tangential-connection term vanishes in the standard
Dirichlet limit; and (iii) neither fact implies a Cotton-type
boundary equation of motion.  A genuine Cotton boundary response
would require a separately justified boundary variational
prescription derived from the original gauge system.  In particular,
we do not add a Pontryagin term or impose self-duality to obtain it.

\section{Pontryagin--Cotton structure and holographic reconstruction}
\subsection{Pontryagin descent and mixed boundary conditions}

For a three-dimensional boundary metric $g_{(0)ij}$, the Cotton
tensor is
\begin{equation}
 C_{ij}=\epsilon_i{}^{kl}\nabla_k
 \left(\mathcal{R}_{lj}-\frac{1}{4}g_{(0)lj}\mathcal{R}\right).
\label{Cotton}
\end{equation}
It is symmetric, traceless, and covariantly conserved.  The metric
variation of the three-dimensional gravitational Chern--Simons
functional is proportional to the Cotton tensor,
\begin{equation}
 \frac{\delta S_{\rm gCS}}{\delta g_{(0)}^{ij}}
 \propto \sqrt{|g_{(0)}|}\,C_{ij}.
\label{CSvariation}
\end{equation}
Consequently this Cotton structure belongs naturally to the descent of
the Pontryagin density $P_4$, Eq.~(\ref{Pontryagin}), rather than to the
Euler density $E_4$ in Eq.~(\ref{Euler}) \cite{DeserJackiwTempleton1982}.

In the holographic description of de Haro \cite{deHaro}, the
renormalized stress tensor and Cotton tensor play roles analogous to
the electric and magnetic components of the gravitational field,
\begin{equation}
 {\cal E}_{ij}\sim\langle T_{ij}\rangle,
 \qquad {\cal B}_{ij}\sim C_{ij}.
\label{EM}
\end{equation}
With a gravitational Chern--Simons boundary term, the variational
principle permits a mixed boundary condition of the schematic form
\begin{equation}
 g_{(3)ij}\propto C_{ij}[g_{(0)}].
\label{deHaroBC}
\end{equation}
The proportionality coefficient depends on the normalization of the
Chern--Simons term and on conventions.  Equation~(\ref{deHaroBC}) is
not a consequence of the bulk Einstein equations alone; those equations
constrain $g_{(3)ij}$ to be transverse and traceless but do not locally
determine it from $g_{(0)ij}$.

The Euler calculation in Sec.~3.1, summarized in
Eq.~(\ref{negativeResult}), shows that the finite Euler boundary form
contains no term linear in $g_3$.
It therefore does not directly yield the mixed
Cotton boundary condition of Eq.~(\ref{deHaroBC}).

\subsection{Two distinct bulk--boundary structures}


The preceding calculation suggests that the Euler and Pontryagin
sectors should be interpreted as two different bulk--boundary
structures,

\begin{equation}
 \boxed{
 \begin{array}{c|c}
 \text{Euler sector} & \text{Pontryagin sector}\\ \hline
 E_4=dQ_E\ (\text{locally}) & P_4=dQ_P\ (\text{locally})\\
 B_3^E=B_3^E(h,K) &
 Q_P=Q_P(\omega)\\
 \text{intrinsic + extrinsic geometry} &
 \text{intrinsic gCS: }Q_{\rm gCS}(\widehat\omega)\\
 \text{global/topological information} &
 \text{Cotton tensor from intrinsic gCS}\\
 \text{no direct linear fixing of }g_{(3)} &
 g_{(3)}\propto C\ \text{under mixed BC}
 \end{array}}
\label{EulerPontryaginTable}
\end{equation}
Here $Q_P$ denotes the full Pontryagin Chern--Simons
form, which may also contain extrinsic-curvature terms,
whereas $Q_{\rm gCS}$ is constructed solely from the
intrinsic boundary connection $\widehat\omega$.
This distinction provides the basis for examining
whether the two sectors can be related by additional
geometric constraints.

A possible unification arises in a self-dual sector.  After symmetry
breaking, define the dual on the four-dimensional Lorentz indices by
\begin{equation}
 ({}^\star R)^{ab}=\frac12\epsilon^{ab}{}_{cd}R^{cd}.
\label{LorentzDual}
\end{equation}
For a (anti-)self-dual Lorentz curvature, Euler and Pontryagin densities
are no longer independent (with the precise sign and factors depending
on Euclidean versus Lorentzian signature and on normalization).  This
raises the possibility that the Euler sector intrinsic to the broken
gauge action and the Pontryagin--Cotton sector relevant to de Haro's
mixed boundary condition are two projections of a common chiral
structure.

This observation also defines a sharper reconstruction problem.  Rather
than assuming the chain $E_4\to C_{ij}\to g_{(3)ij}$, which is not
correct, one should investigate
\begin{equation}
 \boxed{
 \begin{array}{ccc}
 E_4 &\longrightarrow& B_3^E(h,K)\\
 &&\text{global/topological boundary data},\\[2mm]
 P_4 &\longrightarrow& Q_P\\
 && C_{ij}\ \longleftrightarrow\ g_{(3)ij}
 \end{array}}
\label{correctedMap}
\end{equation}
and ask whether a self-duality, regularity, or gauge-theoretic principle
relates the two rows and fixes their relative coefficient.

\subsection{Implications for reconstruction}
\label{secReconstruction}

The Fefferman--Graham expansion separates the boundary source from
normalizable gravitational data:
\begin{equation}
 g_{ij}(z,x)=g_{(0)ij}(x)+z^2g_{(2)ij}(x)
                   +z^3g_{(3)ij}(x)+O(z^4).
 \label{ReconFG}
\end{equation}
For four-dimensional Einstein gravity with negative cosmological
constant, the radial constraints give
\begin{equation}
 g_{(2)ij}=-P_{ij}[g_{(0)}],~~
 \nabla_{(0)}^ig_{(3)ij}=0,~~
 g_{(0)}^{ij}g_{(3)ij}=0.
 \label{ReconConstraints}
\end{equation}
Thus $g_{(2)}$ is fixed locally by $g_{(0)}$, whereas the
transverse-traceless tensor $g_{(3)}$ is not. In particular,
\begin{equation}
 C_{ij}[g_{(0)}]\ \not\Longrightarrow\ g_{(3)ij}.
 \label{ReconNonImplication}
\end{equation}
The holographic stress tensor supplies the missing state-dependent
asymptotic information:
$$
 g_{(3)ij}=\frac{16\pi G}{3\ell^2}\langle T_{ij}\rangle.
 \label{ReconStress}
$$
This relation is the renormalized Dirichlet response established in
Eq.~(\ref{EulerHoloStress}); it is not a consequence of the Cotton
tensor alone.

A concrete reconstruction map can be exhibited in the linearized
transverse-traceless sector about the Poincar\'e patch of AdS$_4$.
For a Fourier mode with spacelike momentum $k^2=q^2>0$, the bulk
Einstein equation and the condition of regularity at $z\to\infty$
select
\begin{equation}
 h_{ij}(z,k)=h_{(0)ij}(k)(1+qz)e^{-qz},
 \label{ReconRegularSolution}
\end{equation}
so that
\begin{equation}
 h_{(3)ij}(k)=\frac{q^3}{3}h_{(0)ij}(k),
 \delta\langle T_{ij}(k)\rangle
 =\frac{\ell^2q^3}{16\pi G}h_{(0)ij}(k).
 \label{ReconResponse}
\end{equation}
For a detailed derivation of Eqs. (\ref{ReconRegularSolution}) and (\ref{ReconResponse}), see Sec.~5.2.2. 
The regularity condition is essential: without it, the second
independent radial solution allows $h_{(3)}$ to vary at fixed
$h_{(0)}$. Eq.~(\ref{ReconResponse}) therefore represents a
particular bulk response map, not a universal local relation
between the two Fefferman--Graham coefficients. 
The linearized boundary Cotton tensor for the same TT source is
\begin{equation}
 C_{ij}^{(1)}(k)
 =\frac{i k^2}{2}\epsilon_i{}^{kl}k_k h_{(0)lj}(k),
 \label{ReconCottonLinear}
\end{equation}
with the orientation convention of Eq.~(\ref{Cotton}).
Unlike Eq.~(\ref{ReconResponse}), this is a local, parity-odd
third-derivative operation on the boundary metric. The two
expressions are not identical, even though both scale as the
third power of boundary momentum.

For example, if one additionally imposes a de Haro-type mixed
boundary condition in the form
\begin{equation}
 h_{(3)ij}=\lambda C_{ij}^{(1)},
 \label{ReconMixed}
\end{equation}
then regularity and the mixed condition must be compatible:
\begin{equation}
 \frac{q^3}{3}h_{(0)ij}-\lambda C_{ij}^{(1)}[h_{(0)}]=0
  \label{ReconCompatibility}
\end{equation}
Here Eq.~(\ref{ReconCompatibility}) is understood on the symmetric
transverse-traceless subspace. It is a restriction on allowed
momenta and polarizations, not an identity for arbitrary sources.
Moreover, Eq.~(\ref{ReconMixed}) is an additional boundary
condition associated with the Pontryagin--Chern--Simons discussion
of Eq.~(\ref{deHaroBC}); it has not been derived from the Euler
sector of the broken conformal gauge action.

The Euler boundary form $B_3^E(h,K)$ provides a distinct
geometrical datum involving intrinsic and extrinsic boundary
geometry. Its local finite term does not linearly fix the
traceless normalizable coefficient $g_{(3)}$. The asymptotic
quantities should therefore be distinguished as
\begin{equation}
 \underbrace{\bigl(g_{(0)ij},g_{(3)ij}\bigr)}_
 {\text{independent FG data}},\qquad
 \underbrace{C_{ij}[g_{(0)}]}_\text{local conformal geometry},
 \qquad
 \underbrace{B_3^E[h,K]}_\text{Euler boundary geometry}.
 \label{ReconData}
\end{equation}
The Schwarzschild--AdS$_4$ example in Sec.~5 gives a particularly
simple check: the conformal boundary metric and its Cotton tensor
are independent of the black-hole mass, whereas $g_{(3)}$ and the
boundary stress tensor retain that mass. These observations
establish the limits of reconstruction from conformal boundary
geometry alone. Reconstructing the complete bulk spacetime further
requires the Einstein evolution equations and appropriate global,
causal, and interior conditions; the asymptotic relations by
themselves do not establish global uniqueness.

\section{Holographic imaging and explicit AdS$_4$ examples}
\label{secImagingExamples}
\subsection{Relation to holographic imaging}
\label{secImaging}

The inverse problem considered by Hashimoto and collaborators
\cite{HashimotoDL,HashimotoRing} provides a useful comparison with the
boundary geometry of the broken conformal gauge theory.  In particular,
the holographic Einstein-ring construction of Ref.~\cite{HashimotoRing}
uses the response of a \emph{scalar operator} to a localized,
time-dependent boundary source.  It should not be identified with the
metric stress-tensor response considered below.

For definiteness, consider a minimally coupled massless probe scalar
$\Phi$ on a fixed asymptotically $\AdS_4$ geometry,
\begin{equation}
 (\Box_{g}-0)\Phi=0,
 \qquad
 \Phi(z,x)=J(x)+z^2\Phi_{(2)}(x)+z^3\Phi_{(3)}(x)+\cdots .
 \label{ImagingScalarFG}
\end{equation}
In the standard quantization, $J$ is the boundary source and the
normalizable coefficient $\Phi_{(3)}$ determines the renormalized
one-point function, up to the conventional normalization and local
contact terms:
\begin{equation}
 \langle {\cal O}(x)\rangle_J
 = {\cal N}_{\Phi}\,\Phi_{(3)}(x)
   +\text{local terms}.
 \label{ImagingScalarResponse}
\end{equation}
The coefficient $\Phi_{(2)}$ is fixed locally by the source and the
boundary metric, whereas $\Phi_{(3)}$ depends on the bulk solution and
its interior boundary condition.  At linear order one may therefore
write the response as
\begin{equation}
 \delta\langle{\cal O}(x)\rangle
 =\int d^3y\,\sqrt{|g_{(0)}(y)|}\,
 G_R(x,y;g_{\mathrm{bulk}})\,\delta J(y),
 \label{ImagingKernel}
\end{equation}
where $G_R$ is the retarded boundary correlator when causal, ingoing
conditions are imposed.  In the black-hole example the bulk metric
enters this kernel through the propagation of the probe field.  The
imaging procedure processes such a boundary response to extract
information about the bulk geometry \cite{HashimotoRing}.

The gravitational reconstruction problem has a parallel, but distinct,
source--response pair.  For Dirichlet boundary conditions the source is
$g_{(0)ij}$ and the state-dependent response is
$$ \langle T_{ij}\rangle
 =\frac{3\ell^2}{16\pi G}\,g_{(3)ij},
 $$
in agreement with the renormalized Dirichlet relation derived in Sec.~3.2.  Thus the two schematic maps are
\begin{equation}
 \begin{aligned}
 \text{probe imaging:}\quad
 &J\ \longrightarrow\ \langle{\cal O}\rangle
     \ \longrightarrow\ \text{information on }g_{\mathrm{bulk}},\\
 \text{metric response:}\quad
 &g_{(0)ij}\ \longrightarrow\ g_{(3)ij}
     \ \longrightarrow\ \langle T_{ij}\rangle.
 \end{aligned}
 \label{ImagingTwoMaps}
\end{equation}
The second arrow in the metric line is the holographic dictionary;
the first arrow is \emph{not} determined by the near-boundary Einstein
constraints alone.  It requires a choice of bulk state and suitable
regularity, causal, or interior conditions.  The same distinction
between source data and normalizable response appears in the probe
problem.

The gauge-theoretic characteristic forms organize different boundary
information:
\begin{equation}
 E_4\longrightarrow B_3^E(h,K),
 \qquad
 P_4\longrightarrow Q_P\longrightarrow C_{ij}[g_{(0)}].
 \label{ImagingGeometricMaps}
\end{equation}
Neither $B_3^E$ nor $C_{ij}[g_{(0)}]$ is, by itself, the scalar
response kernel $G_R$.  Furthermore, the Cotton tensor depends locally
on the boundary metric and cannot generally replace the independent
normalizable gravitational datum $g_{(3)ij}$.  A Cotton--stress relation
requires an additional mixed boundary condition, as discussed in
Sec.~4; it is not implied by the Euler boundary form.

We now examine two explicit examples illustrating the distinction between boundary conformal geometry and normalizable bulk data. The Schwarzschild--AdS$_4$ family shows that different bulk masses can share the same boundary Cotton tensor, whereas regular linearized perturbations of Poincaré AdS$_4$ provide an explicit momentum-dependent relation between the boundary source and the normalizable gravitational response.

\subsection{Explicit asymptotically AdS geometries}\label{secExamples}
\subsubsection{Pure AdS$_4$ and Schwarzschild--AdS$_4$}
Before considering a general reconstruction problem, it is useful
to test the proposal for simple asymptotically AdS geometries.

For a conformally flat boundary,
\begin{equation}
 C_{ij}=0.
\end{equation}

Pure AdS$_4$ provides the reference solution with
$C_{ij}=0$ and $g_{(3)ij}=0$.
In contrast, Schwarzschild--AdS$_4$ has the same
conformal boundary geometry but a nonvanishing
normalizable mode $g_{(3)ij}$ when $M\neq 0$.
This example is trivial but important: it shows immediately that
$C_{ij}=0$ by itself cannot distinguish all possible bulk states
with conformally flat boundary geometry.
Thus regularity and/or stress-tensor data must enter the complete
reconstruction problem.

A concrete distinction between boundary conformal geometry and
normalizable gravitational data is provided by Schwarzschild--AdS$_4$.
This is also the black-hole background relevant to the imaging
proposal of Ref.~\cite{HashimotoRing}. Consider
\begin{equation}
 ds^2=-f(r)dt^2+\frac{dr^2}{f(r)}+r^2d\Omega_2^2,
 \qquad f(r)=1+\frac{r^2}{\ell^2}-\frac{2GM}{r}.
 \label{SAdSmetric}
\end{equation}
We introduce dimensionless quantities
\begin{equation}
 R=\frac r\ell,\qquad \tau=\frac t\ell,\qquad
 m=\frac{GM}{\ell},
\end{equation}
and choose the dimensionless Fefferman--Graham coordinate $z$
so that $R\sim z^{-1}$ at the conformal boundary. The radial
coordinate transformation obeys
\begin{equation}
 \frac{dR}{dz}=-\frac1z\sqrt{1+R^2-\frac{2m}{R}}.
 \label{SAdSradial}
\end{equation}
Substituting $R=z^{-1}+az+bz^2+O(z^3)$ gives
\begin{equation}
 a=-\frac14,\qquad b=\frac m3,\qquad
 R(z)=\frac1z-\frac z4+\frac m3 z^2+O(z^3).
 \label{SAdSradialExpansion}
\end{equation}
The metric consequently takes the Fefferman--Graham form
\begin{equation}
 ds^2=\frac{\ell^2}{z^2}
 \left[dz^2+g_{ij}(z,x)dx^idx^j\right],
 \qquad x^i=(\tau,\theta,\phi),
\end{equation}
with $g_{ij}=g_{(0)ij}+z^2g_{(2)ij}+z^3g_{(3)ij}+O(z^4)$.
Writing $\gamma_{AB}$ for the metric of the unit two-sphere,
we obtain
\begin{align}
 g_{\tau\tau}
 &=-z^2\left(1+R^2-\frac{2m}{R}\right)
 =-1-\frac{z^2}{2}+\frac{4m}{3}z^3+O(z^4),\\
 g_{AB}
 &=z^2R^2\gamma_{AB}
 =\left(1-\frac{z^2}{2}+\frac{2m}{3}z^3\right)
 \gamma_{AB}+O(z^4).
\end{align}
Thus
\begin{align}
 g_{(0)ij}dx^idx^j&=-d\tau^2+d\Omega_2^2,
 \label{SAdSg0}\\
 g_{(2)ij}dx^idx^j&=-\frac12(d\tau^2+d\Omega_2^2),
 \label{SAdSg2}\\
 g_{(3)ij}dx^idx^j&=\frac{2m}{3}
 (2d\tau^2+d\Omega_2^2).
 \label{SAdSg3}
\end{align}
The mass appears first in $g_{(3)ij}$, which satisfies
$g_{(0)}^{ij}g_{(3)ij}=0$ and
$\nabla_{(0)}^ig_{(3)ij}=0$.

The boundary metric $\mathbb R\times S^2$ has Ricci tensor
$\mathcal{R}_{\tau\tau}=0$, $\mathcal{R}_{AB}=\gamma_{AB}$, and scalar curvature
$\mathcal{R}=2$. Its three-dimensional Schouten tensor is
\begin{equation}
 P_{\tau\tau}=\frac12,\qquad
 P_{AB}=\frac12\gamma_{AB},\qquad
 \nabla_{(0)k}P_{ij}=0.
\end{equation}
Hence both the three-index Cotton tensor and its two-index dual
vanish identically:
\begin{equation}
 C_{ijk}[g_{(0)}]=0,\qquad C_{ij}[g_{(0)}]=0,
 \label{SAdSCotton}
\end{equation}
independently of $M$.

In contrast, the renormalized holographic stress tensor,
Eq.~(\ref{EulerHoloStress}), gives
\begin{equation}
 \langle T^i{}_j\rangle_{g_{(0)}}
 =\frac{M\ell}{8\pi}\operatorname{diag}(-2,1,1).
 \label{SAdSstress}
\end{equation}
This expression uses the dimensionless boundary representative
$-d\tau^2+d\Omega_2^2$. In the physical conformal frame
$-dt^2+\ell^2d\Omega_2^2$, the mixed components become
\begin{equation}
 \langle T^i{}_j\rangle_{\mathrm{phys}}
 =\frac{M}{8\pi\ell^2}\operatorname{diag}(-2,1,1),
\end{equation}
and the conserved energy is
\begin{equation}
 E=-\int_{S^2}d^2x\sqrt{\sigma}\,
 \langle T^t{}_t\rangle_{\mathrm{phys}}=M.
 \label{SAdSenergy}
\end{equation}
Thus the same conformal boundary geometry, with $C_{ij}=0$,
corresponds to different values of $g_{(3)ij}$ and different
bulk masses. In particular,
\begin{equation}
 C_{ij}=0\quad\not\Longrightarrow\quad g_{(3)ij}=0.
 \label{SAdSnonimplication}
\end{equation}
This directly illustrates the non-implication discussed in
Sec.~4 without imposing a mixed boundary condition.

The Euler boundary sector contains additional extrinsic data.
For the outward-normal convention of Sec.~3, the Fefferman--Graham expansion
of the mixed extrinsic curvature is
\begin{equation}
 K^i{}_j=\frac1\ell\delta^i_j
 +\frac{z^2}{\ell}P^i{}_j
 -\frac{3z^3}{2\ell}g_{(3)}{}^i{}_j+O(z^4).
\end{equation}
In this example it becomes
\begin{equation}
 K^i{}_j=\frac1\ell\delta^i_j
 +\frac{z^2}{2\ell}\operatorname{diag}(-1,1,1)
 -\frac{mz^3}{\ell}\operatorname{diag}(-2,1,1)
 +O(z^4).
 \label{SAdSextrinsic}
\end{equation}
Although the mass-dependent normalizable mode occurs in
$K^i{}_j$, its term linear in $g_{(3)}$ in the finite local
Euler boundary density cancels by
$\operatorname{tr}g_{(3)}=0$, as established in Sec.~3.
This local asymptotic cancellation must not be confused with
mass independence of the integrated Euler invariant, which
can involve global and horizon contributions.

Schwarzschild--AdS$_4$ therefore exhibits explicitly the
separation of the boundary Cotton tensor, the state-dependent
holographic stress tensor, and the Euler intrinsic--extrinsic
boundary geometry. Computing the dynamical imaging response
itself would require solving a perturbation equation on this
black-hole background, rather than inferring that response
from the Cotton tensor alone.

\subsubsection{Perturbed AdS$_4$}
\label{secPerturbedAdS}

We next consider a linearized gravitational perturbation of the
Poincar\'e patch of AdS$_4$.  This example distinguishes the
nonlocal response determined by the bulk equations and a regularity
condition from the local Cotton tensor of the boundary source.  We
work in Fefferman--Graham gauge,
\begin{equation}
 ds^2=\frac{\ell^2}{z^2}\left[dz^2+
 (\eta_{ij}+h_{ij}(z,x))dx^i dx^j\right],\qquad
 \eta_{ij}=\operatorname{diag}(-1,1,1),
 \label{PertFG}
\end{equation}
and restrict to transverse-traceless perturbations,
\begin{equation}
 \partial^i h_{ij}=0,\qquad \eta^{ij}h_{ij}=0.
 \label{PertTT}
\end{equation}
Writing $h_{ij}(z,x)=e^{ik\cdot x}H_{ij}(z,k)$, the linearized
Einstein equations reduce to
\begin{equation}
 \left(\partial_z^2-\frac{2}{z}\partial_z-k^2\right)
 H_{ij}(z,k)=0,\qquad k^2=\eta^{ij}k_i k_j.
 \label{PertODE}
\end{equation}
For spacelike momentum $k^2=q^2>0$, the general solution is
\begin{equation}
 H_{ij}=A_{ij}(1+qz)e^{-qz}
       +B_{ij}(1-qz)e^{qz}.
 \label{PertGeneral}
\end{equation}
Selecting the solution regular as $z\to\infty$ sets $B_{ij}=0$.
The near-boundary expansion is then
\begin{align}
 H_{ij}(z,k)
 &=h_{(0)ij}(k)(1+qz)e^{-qz}\notag\\
 &=h_{(0)ij}(k)\left[1-\frac{q^2z^2}{2}
       +\frac{q^3z^3}{3}+O(z^4)\right],
 \label{PertRegular}
\end{align}
so that
\begin{equation}
 h_{(2)ij}=-\frac{q^2}{2}h_{(0)ij},\qquad
 h_{(3)ij}=\frac{q^3}{3}h_{(0)ij}.
 \label{PertFGcoeff}
\end{equation}
The second relation is a consequence of the specified interior
regularity condition; it is not a local Fefferman--Graham constraint valid for
arbitrary states.  Using the holographic stress tensor in
Eq.~(\ref{EulerHoloStress}) gives the corresponding linear response,
\begin{equation}
 \delta\langle T_{ij}(k)\rangle
 =\frac{3\ell^2}{16\pi G}h_{(3)ij}(k)
 =\frac{\ell^2}{16\pi G}(k^2)^{3/2}h_{(0)ij}(k).
 \label{PertStress}
\end{equation}
This expression is derived in the spacelike-momentum sector;
retarded Lorentzian response requires the corresponding causal
analytic continuation and horizon/interior prescription.

For comparison, define the boundary Cotton tensor by Eq.~(\ref{Cotton}).

The overall sign depends on the orientation convention.  At first
order about the flat boundary metric, Eq.~(\ref{PertTT}) implies
\begin{equation}
 \mathcal{R}_{ij}^{(1)}=-\frac12\Box h_{(0)ij},\qquad
 \mathcal{R}^{(1)}=0,
\end{equation}
and hence
\begin{align}
 C_{ij}^{(1)}
 &=-\frac12\epsilon_i{}^{kl}\partial_k\Box
 h_{(0)lj},\notag\\
 C_{ij}^{(1)}(k)
 &=\frac{i}{2}k^2\epsilon_i{}^{kl}k_k
 h_{(0)lj}(k).
 \label{PertCottonLinear}
\end{align}
Thus the stress response and Cotton tensor are both third order in
boundary momentum but act differently on the transverse-traceless
polarizations: the former is proportional to $(k^2)^{3/2}$ times
the source, whereas the latter includes the parity-odd operator
$\epsilon_i{}^{kl}k_k$.  In particular, the two tensors are not
identical for a generic boundary perturbation.

We can test directly whether the Euler boundary three-form adds a
finite contribution linear in the normalizable mode.  Write the
mixed extrinsic curvature, consistently with Eq.~(\ref{Kmixed}), as
\begin{equation}
 K^i{}_j=\ell^{-1}\delta^i{}_j
 +\ell^{-1}z^2 P^i{}_j
 -\frac{3z^3}{2\ell}h_{(3)}{}^i{}_j+O(z^4).
 \label{PertK}
\end{equation}
The $z^3$ perturbation in Eq.~(\ref{PertFGcoeff}) is
\begin{equation}
 \left.\delta K^i{}_j\right|_{z^3}
 =-\frac{q^3z^3}{2\ell}h_{(0)}{}^i{}_j,
 \qquad
 \left.\delta K^i{}_i\right|_{z^3}=0.
 \label{PertKTT}
\end{equation}
For the cubic extrinsic-curvature term in $B_3^E$, use the
three-dimensional identity already established in
Eq.~(\ref{detK}).
At first order in the normalizable perturbation,
\begin{align}
 \delta_{(3)}\det K
 &=\det K\,\operatorname{Tr}(K^{-1}\delta_{(3)}K)
 \notag\\
 &=\frac{1}{\ell^2}\operatorname{Tr}(\delta_{(3)}K)
    +O(z^5)
 =O(z^5),
 \label{PertDetCancel}
\end{align}
where the would-be $z^3$ term vanishes because
$\operatorname{Tr}h_{(3)}=0$.  Since
$\sqrt{|h|}=\ell^3z^{-3}\sqrt{|g_{(0)}|}[1+O(z^2)]$,
the cubic sector has no finite $z^0$ density linear in
$h_{(3)}$.  The remaining intrinsic-curvature terms are of the
form $\sqrt{|h|}(K\mathcal R-2K^{ij}\mathcal R_{ij})$.
Their part linear in $h_{(3)}$ is at least
\begin{equation}
 \sqrt{|h|}\,\mathcal R^i{}_j[h]\,\delta_{(3)}K^j{}_i
 \sim z^{-3}\,z^2\,z^3=O(z^2),
 \label{PertEulerScaling}
\end{equation}
and likewise has no finite $z^0$ contribution.  Thus,
\begin{equation}
 \boxed{\left.B_3^E\right|_{z^0,\,\mathrm{linear\ in}\ h_{(3)}}=0.}
 \label{PertEulerResult}
\end{equation}
This is a statement about the local asymptotic density, not about
the full on-shell variation of the Euler-completed action.  The
latter also involves the connection variations discussed in
Sec.~\ref{secEulerVariation}, and yields the standard renormalized
stress tensor under the prescribed boundary conditions.

The explicit comparison is therefore
\begin{equation}
 \boxed{\begin{aligned}
 h_{(3)ij}(k)&=\frac{q^3}{3}h_{(0)ij}(k),\\
 \delta\langle T_{ij}(k)\rangle
   &=\frac{\ell^2q^3}{16\pi G}h_{(0)ij}(k),\\
 C_{ij}^{(1)}(k)&=\frac{iq^2}{2}
    \epsilon_i{}^{kl}k_kh_{(0)lj}(k).
 \end{aligned}}
 \label{PertComparison}
\end{equation}
No equality between $C_{ij}^{(1)}$ and $h_{(3)ij}$ follows from
the Euler sector alone.  A relation of that type requires an
additional mixed boundary condition, a self-duality restriction,
or some other independent dynamical input.  The present calculation
thus makes precise, at linear order, the distinction between the
boundary response and the conformal boundary geometry.

\section{Discussion and conclusions}

The present analysis clarifies the distinct roles of the
Euler and Pontryagin sectors in the holographic
interpretation of gravity arising from broken
$SO(4,2)$ gauge symmetry. Although both characteristic
densities are locally total derivatives, their
boundary structures are different:
$$
 E_4=\epsilon_{abcd}R^{ab}\wedge R^{cd},
 \qquad
 P_4=R^{ab}\wedge R_{ab}.
$$
The Euler boundary transgression involves both
intrinsic and extrinsic geometry. The Pontryagin
boundary form may also involve extrinsic curvature,
while its intrinsic gravitational Chern--Simons
counterpart has a Cotton metric variation.

Our explicit Fefferman--Graham calculation shows that
the finite Euler boundary density contains no term
linear in the normalizable coefficient $g_{(3)ij}$.
The cubic extrinsic-curvature contribution vanishes
because $\mathrm{Tr}\,g_{(3)}=0$, while the corresponding
intrinsic-curvature contributions vanish in the
conformal-boundary limit. This result does not imply
the absence of a holographic stress-tensor response
from the complete Euler-completed action. Rather,
it shows that the Euler boundary form alone does not
generate the local mixed condition
$g_{(3)ij}\propto C_{ij}[g_{(0)}]$.

The Codazzi analysis provides a complementary result:
the boundary Cotton tensor appears in the asymptotic
mixed curvature, but its contribution through the
tangential-connection variation vanishes under
standard Dirichlet boundary conditions. Thus the
kinematical appearance of the Cotton tensor must be
distinguished from a dynamical Cotton boundary
response.

These results suggest two complementary structures
of bulk--boundary information:
\begin{equation}
\boxed{\begin{gathered}
 \text{Euler: topological/global bulk--boundary relation}\\[3pt]
 \oplus\\[3pt]
 \text{Pontryagin: local Cotton/electric--magnetic relation}
\end{gathered}}
\end{equation}
In holographic reconstruction, state-dependent
information is encoded in boundary response functions,
while the gauge-theoretic characteristic sectors
provide additional geometric and topological
structure. A possible connection between the two
sectors through self-duality or chirality of the
Lorentz curvature remains an open question. Establishing such a connection, including its normalization and signature dependence, may clarify
whether Cotton-type mixed boundary conditions can
emerge from the broken gauge theory itself.
A further connection may be found in the Ryu--Takayanagi prescription for holographic entanglement entropy \cite{RT}. In AdS$_4$/CFT$_3$, the entanglement entropy of a boundary region is related to the area of a bulk codimension-two extremal surface. The four-dimensional Euler term can also contribute a topological term associated with the Euler characteristic of this surface, provided the appropriate boundary contributions are included. This is distinct from the finite Euler boundary density and its Dirichlet variation examined in the present work. A direct derivation of the entanglement contribution from the Euler sector of the broken conformal gauge action, including its fixed normalization, remains an interesting problem for future investigation.


\end{document}